# Anisotropic London Penetration Depth and Superfluid Density in Single Crystals of Iron-based Pnictide Superconductors

R. Prozorov, M. A. Tanatar, R. T. Gordon, C. Martin, H. Kim, V. G. Kogan,
N. Ni, M. E. Tillman, S. L. Bud'ko and P. C. Canfield

Ames Laboratory and Department of Physics & Astronomy,
Iowa State University, Ames, IA 50011

**Abstract**

In- and out-of–plane magnetic penetration depths were measured in three iron-based pnictide superconducting systems. The "122" system was represented by electron-doped Ba(Fe$_{1-x}$Co$_x$)$_2$As$_2$ with the doping through the whole phase diagram with x ≈ 0.038, 0.047, 0.058, 0.074 and 0.10 ($T_c$ ranged from 13 to 24 K) and by hole-doped (Ba$_{1-x}$K$_x$)Fe$_2$As$_2$ with doping close to optimal, with measured $x \approx 0.45$ ($T_c \approx 28$ K) and an underdoped sample with $x \approx 0.15$ ($T_c \approx 19$ K). The "1111" system was represented by single crystals of NdFeAs(O$_{1-x}$F$_x$) with nominal $x = 0.1$ ($T_c \approx 43$ K). All studied samples of both 122 systems show a robust power-law behavior, $\lambda(T) \propto T^n$, with the sample-dependent exponent $n = 2 - 2.5$, which is indicative of unconventional pairing. This scenario could be possible either through scattering in a $s_{\pm}$ state or due to nodes in the superconducting gap. In the Nd-1111 system, the interpretation of the results is complicated by magnetism of the rare-earth ions. For all three systems, the anisotropy ratio, $\gamma_\lambda \equiv \lambda_c / \lambda_{ab}$, was found to decrease with increasing temperature, whereas the anisotropy of the coherence lengths, $\gamma_\xi \equiv \xi_{ab}/\xi_c = H_{c2}^{\perp c}/H_{c2}^{\|c}$, has been found to increase (both opposite to the trend in two-band MgB$_2$). The overall anisotropy of the pnictide superconductors is small, in fact much smaller than that of the cuprates (except YBa$_2$Cu$_3$O$_{7-x}$ (YBCO)). The 1111 system is about two times more anisotropic than the 122 system. Our data and analysis suggest that the iron-based pnictides are complex superconductors in which a multiband three-dimensional electronic structure and strong magnetic fluctuations play important roles.

# Introduction

Precision measurements of the temperature-dependent London penetration depth are among the most useful tools to probe low-energy quasiparticles in superconductors [1]. The information obtained can be used to distinguish between different possible superconducting gap structures on the Fermi surface and ultimately shed light on the superconducting pairing mechanism. While determination of the type of superconductivity requires phase-sensitive experiments, such experiments are very difficult and so far have been performed only on a few superconducting systems [2, 3]. Other direct probes that access the density of states are angle-resolved photoemission (ARPES) [4] and tunneling [5]. These techniques are very useful and important. However, extreme sensitivity to the state of the surface (only the topmost surface layer is probed) and limited directional resolution make them difficult to use if the Fermi surface is significantly three-dimensional. The magnetic penetration depth on the other hand is a bulk probe that can be coupled to all components of the superfluid response tensor [6], thus probing all directions in the reciprocal space. In fact, it was microwave measurements of the penetration depth that first established the d-wave nature of superconductivity in the high-$T_c$ cuprates [7].

Superconductivity in LaFeAsO$_{1-x}$F$_x$ with $T_c \approx 23$ K was discovered less than a year ago [8] and shortly after materials with other rare earth (R) elements were synthesized, raising the critical temperatures above 50 K in RFeAsO$_x$F$_y$ (R=Nd,Sm,Pr) [9-12]. This is significantly higher than the highest $T_c$ reported in conventional s-wave superconductors and not surprisingly these discoveries have attracted a lot of attention. The compounds with the original structure, RFeAsO$_x$F$_y$, are now frequently abbreviated as "1111". In superconducting samples, x is always less than 1, while y can be zero (fluorine-free 1111 pnictides). Later in the year, superconductors of a different crystal structure were discovered, based on the parent AFe$_2$As$_2$ (abbreviated as "122", here A is an alkaline earth element) [13]. In the 122 system, either the A or Fe sites can be doped to achieve superconductivity with holes or electrons as carriers. Unlike the oxygen-based 1111 system, large single crystals of the oxygen-free 122 system have become readily available [14-17]. For several obvious reasons, experiments aimed at understanding the superconducting gap symmetry can be done only on single crystals. First, they allow for the study of the anisotropy of the response, which is linked to the structural, electronic and superconducting anisotropy of the material. This is especially important for superconductors with anisotropic gaps and gaps with nodes, having reduced symmetry as compared to the lattice [18]. Second, the interpretation of the results on polycrystalline materials is often hindered by various extrinsic factors, such as grain boundaries, morphological defects and uncertainty in the sample volume and internal structure. For most electromagnetic measurements, these factors could lead to incorrect conclusions regarding the superconducting pairing.

Many experiments have been performed to determine the pairing symmetry of the pnictide superconductors and their review is beyond the scope of this paper. This special volume of Physica C covers most of the different techniques used in these studies. We will only comment on the penetration depth measurements in single crystals. Similar to the case of YBCO [7], the first report came from the microwave cavity perturbation technique that found fully-gapped superconductivity in $PrFeAsO_{1-y}$ [19]. That work was followed by the radio-frequency measurements of $\lambda(T)$ in single crystals of $SmFeAs(O_{1-x}F_x)$ [20] and $NdFeAs(O_{1-x}F_x)$ [21]. These have found an exponential low-temperature behavior consistent with a fully gapped state. However, the crystals studied were very small (< 50 µm), resulting in a low signal-to-noise ratio in the data. Single crystals of the 1111 system are still small and unfortunately single crystals of the nonmagnetic La-1111 have yet to be synthesized. Other rare-earth (Nd, Pr, Sm,) systems are known to be magnetic [22] and it is quite possible that this "exponential" behavior has the same origin as in the electron-doped cuprates, such as $Nd_{1.85}Ce_{0.15}CuO_{4-x}$ (NCCO), that for ten years were believed to be s-wave superconductors until the influence of the $Nd^{3+}$ magnetism was resolved [23-25]. There are several bulk measurements, such as NMR, that show the absence of a coherence peak and a $T^3$ behavior of the $1/T_1$ relaxation rate, indicating that the superconductivity is not fully gapped [26]. Field-dependent specific heat measurements were interpreted in terms of gap with nodes [27]. On the other hand, ARPES measurements indicate fully gapped Fermi surfaces (at least in the $k_z = 0$ plane), albeit that not all five sheets are observed. (We also note that if we take values of $\Delta(0)/T_c$ reported by ARPES and calculate the superfluid density, the result will be very far from the experimental observations). At this point we would like to leave the question of the order parameter symmetry in the 1111 system wide open and simply state that additional experiments are needed.

The situation with the 122 system has been found to be even more interesting, thanks to the availability of large high quality single crystals. Microwave cavity perturbation measurements were used to study three hole-doped $(Ba_{1-x}K_x)Fe_2As_2$ (BaK-122) crystals with various amount of disorder and they report a fully-gapped Fermi surface with two distinct gaps [28]. Our radio-frequency measurements performed on BaK-122 crystals from different sources and grown from different fluxes indicate a robust power-law behavior of the penetration depth, $\Delta\lambda \propto T^n$ [29]. μSR measurements performed on samples with x=0.5 have found a close to linear variation of the superfluid density [30]. Furthermore, studies of the electron-doped $Ba(Fe_{1-x}Co_x)_2As_2$ (FeCo-122) also consistently show a power-law behavior across the broad range of cobalt concentrations with the exponent $n$ in the range of 2-2.5 [31, 32]. Similar results in a 122 system were obtained by Bristol group on crystals from yet different source [17, 33].

Even more surprising is the recent report of a robust *T*-linear behavior of the penetration depth in very clean LaFePO single crystals [34]. Although not containing arsenic and having a considerably lower $T_c \approx 6$ K, this material is very close to the arsenic-containing 1111 system with regard to its electronic structure.

It is indeed very puzzling that these precision measurements performed on high quality crystals provide such different information. Obviously, more measurements and controlled doping- and pressure-dependent studies are needed.

Another important issue is the anisotropy in these materials. Initial reports have provided various estimates, some of which are very large. The current picture is that none of the pnictide superconductors possess considerable values of anisotropy. Within the Ginzburg-Landau theory, the anisotropy of the penetration depths, $\gamma_\lambda(T) = \lambda_c/\lambda_{ab}$, of the upper critical fields, $\gamma_\xi(T) = \xi_{ab}/\xi_c = H_{c2}^{\|ab}/H_{c2}^{\|c}$ and of the normal-state resistivities, $\gamma_\rho(T) = \rho_{ab}/\rho_c$, are related at the transition temperature, $T_c$, via $\gamma_\lambda(T_c) = \gamma_\xi(T_c) = \sqrt{\gamma_\rho(T_c)}$ (see Ref.[26] in Ref.[35]). It has been found that at $T_c$ the 1111 system shows $\gamma_\lambda(T_c) \approx 4$ and increases to about 18 at 1 K. Similar behavior was inferred from measurements of the first critical field in Pr-1111 crystals with a miniature Hall probe [36] as well as in Nd- and Sm-1111 crystals from torque measurements [37-39]. The 122 system is even less anisotropic, - $\gamma_\lambda$ is about 2 at $T_c$ and it increases to about 6 at 1 K. Such low anisotropies make the pnictides very distinct from the high-$T_c$ cuprates (except for YBCO which is also only moderately anisotropic).

In this paper we review our most up to date results on the anisotropic London penetration depth in single crystals of the 122 system for both the hole doped $Ba_{1-x}K_xFe_2As_2$ (measured x ≈ 0.45 and 0.15) and the electron doped $Ba(Fe_{1-x}Co_x)_2As_2$ ( measured x ≈ 0.038, 0.047, 0.058, 0.074 and 0.10) as well as the Nd-based member of the 1111 system, $NdFeAs(O_{1-x}F_x)$, nominal x≈0.1. In all 122 crystals, we find a power-law behavior of the penetration depth, $\lambda(T) \sim T^n$, down to the lowest temperature corresponding to $T/T_c \approx 0.02$ with the exponent $n$ close to 2, but somewhat larger. In the Nd-1111 system, the situation is less clear due to much smaller crystal sizes and the influence of the magnetism of the $Nd^{3+}$ ions. For all systems we show the experimentally determined anisotropy of the penetration depth, $\gamma_\lambda(T)$, which increases with decreasing temperature. Taking into account the opposite trend in the temperature dependence of $\gamma_\xi(T)$ (e.g., in the Co-122 system, showing a weak decrease from about 2 at $T_c$ to about 1 at low temperatures [15]) the overall behaviors of the temperature-dependent anisotropies are literally opposite to those observed in MgB$_2$ [40, 41].

## Experimental

### *Fe-based pnictide single crystals*

Single crystals of the 122 system were grown at ambient pressure using a flux-growth technique. Single crystals of $Ba(Fe_{1-x}Co_x)_2As_2$ ("FeCo-122", x ≈0.038, 0.047, 0.058, 0.074 and 0.10, $T_c$ ranged from 13 to 24 K) were grown out of self flux [15]. Single crystals of (Ba$_{1-x}$K$_x$)Fe$_2$As$_2$ ("BaK-122", $x=0.45$ ($T_c \approx 28$ K) and $x=0.15$ ($T_c \approx 19$ K)) were grown out of a tin flux [14]. The actual cobalt and potassium concentrations were determined by wavelength dispersive x-ray spectroscopy in the electron probe microanalyzer of a JEOL JXA-8200 Superprobe. The quality of the samples was checked with magneto-optics. Figure 1 (a) shows uniform trapped magnetic flux in a FeCo-122 crystal with x=0.074 (also see Ref.[42]).

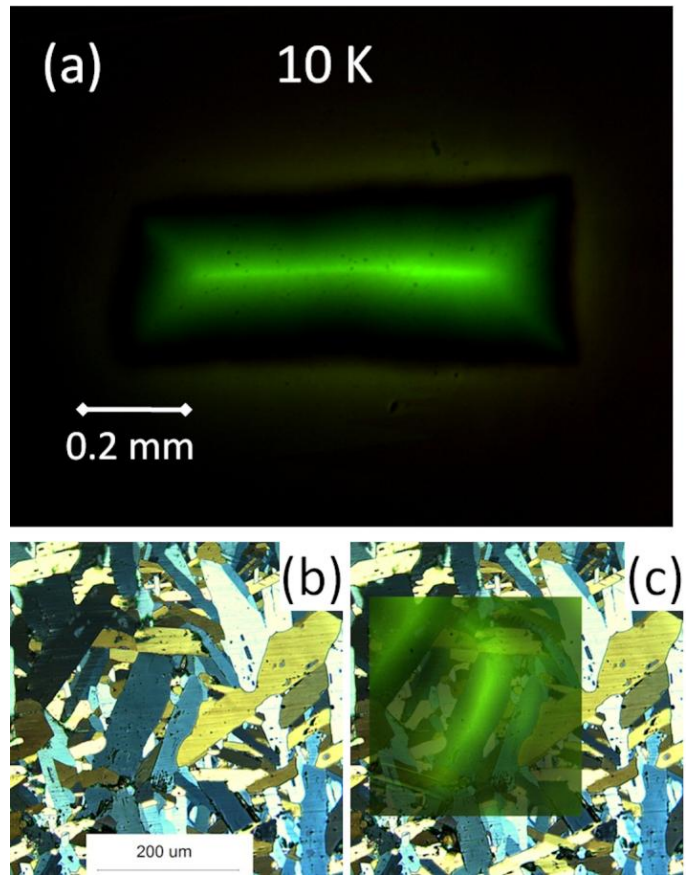

Figure 1. (a) Magneto-optical image of trapped flux in a Ba(Fe$_{1-x}$Co$_x$)$_2$As$_2$ (x=0.074) single crystal; (b) polarized-light image of a polished surface of a NdFeAs(O$_{0.9}$F$_{0.1}$) pellet; (c) superimposed image of the trapped flux clearly showing ~200 nm long superconducting crystallite.

Single crystals of the 1111 system were obtained by using a high-pressure synthesis technique. Samples with the nominal composition of NdFeAs(O$_{0.9}$F$_{0.1}$) (Nd-1111) were grown in a cubic multianvil apparatus from Rockland Research Corporation at a maximum pressure of about 3.3 GPa. The synthesis yields dense polycrystalline NdFeAsO$_{0.9}$F$_{0.1}$ samples that contain plate-like single crystals as large as 500 μm. The extraction of these crystals is possible with the help of magneto-optical imaging that allows for the detection of the best superconducting regions [21, 43]. This is shown in Figure 1 (b) and (c). The first panel is the polarized-light image of a polished surface of a 5 mm diameter pellet obtained using high-pressure synthesis. Sensitivity to the orientation of the light polarization plane with respect to the crystal structure (producing different colors) serves as an additional indication that we are dealing with well ordered crystallites. Tetragonal symmetry of the unit cell suggests that the crystals should grow as plates with the $ab-$ plane being the extended surface and the c - axis along the shortest dimension. We therefore expect that, statistically, crystallites with the largest areas are those whose c-axis is perpendicular to the image plane. This was directly confirmed by x-ray diffraction after these crystallites were extracted. Figure 1 (c) shows an overlay of a magneto-optical image that clearly reveals one relatively large good single crystal. In our definition of "good" we look for a spatially uniform trapped magnetic flux and we also check for the uniformity of Meissner screening [43]. Apparently, only some crystallites are superconducting and this is why magneto-optical imaging was very important for the identification of good single crystals. The tunnel-diode resonator data reported in this paper were obtained on one of the largest crystals available for a 1111 system with dimensions of $120\times650\times80$ μm$^3$.

### *Magneto-optical imaging*

Magneto-optical (MO) imaging was performed in a $^4$He optical flow-type cryostat using Faraday rotation of polarized light in a Bi-doped iron-garnet film having an in-plane magnetization. The spatial resolution of this technique is about 3 μm with sensitivity to magnetic fields of about 1 G. The temporal resolution is about 30 msec [42, 44]. In a 1111 system (see Figure 1 (b) and (c)), obtained using high-pressure synthesis, magneto-optical imaging was used to identify the most homogeneous Meissner screening and trapped flux within the samples and in our opinion provides a good evaluation of the quality of a superconductor at least down to the mesoscopic scale of about 3 μm [43]. For the 122 system, slabs with sizes of $\sim 1\times1\times0.2$ mm$^3$ having mirror-like surfaces, were cleaved with a razor blade from larger crystals. The obtained samples were subsequently imaged with a magneto-optical setup to ensure their quality, Figure 1 (a).

### *Tunnel-diode resonator (TDR) technique*

The penetration depth $\lambda(T)$ was measured by placing the sample inside the coil of a tank circuit biased by a tunnel-diode and resonating at a radio-frequency with a self-resonant frequency of $f_0 \approx 14$ MHz. The excitation ac magnetic field $H_{ac} \approx 10$ mOe is much less than the lower critical field $H_{c1} \approx 100$ Oe, as estimated from low-field magnetization measurements. Therefore, in zero dc applied fields the sample is in the Meissner state. Upon insertion of a sample with magnetic susceptibility $\chi$ into the coil, the resonant frequency change, $\delta f(T) = f_0 - f(T)$, is given by [1, 45, 46],

$$\left. \begin{array}{c} 2\pi f(T) = 1/\sqrt{L(T)C} \\ \delta f_{max} / f_0 \leq 10^{-3} \end{array} \right\} \Rightarrow \frac{\delta f(T)}{f_0} \simeq -\frac{\Delta L(T)}{2L_0} \approx -\frac{V_{sample}}{2V_{coil}} 4\pi\chi(T) \qquad (1)$$

Here $f(T)$ is the actual resonant frequency, $L(T)$ is the inductance of the coil containing the sample, $f_0$ is the resonant frequency of the empty resonator and $\delta f_{max}$ represents the maximum frequency shift we encounter in our experiments. $V_{sample}$ and $V_{coil}$ are the volumes of the sample and of the coil, respectively, and $\chi(T)$ is the *dynamic* magnetic susceptibility of the *entire* sample defined as $\chi = dM/dH$ (this also includes demagnetization effects). With a local magnetic permeability $\mu$ (e.g., due to magnetic $Nd^{3+}$ ions) the global dynamic susceptibility of a slab can be written as

$$4\pi\chi \simeq \mu \frac{\lambda(\mu)}{R} \tanh\frac{R}{\lambda(\mu)} - 1 \qquad (2)$$

and in the case of dilute magnetic impurities, $\lambda(\mu) = \lambda_L / \sqrt{\mu}$, where $\lambda_L$ is the London penetration depth of the "non-magnetic" material [47, 48]. Therefore, by measuring the frequency shift due to the influence of the sample, one can measure the change in the London penetration depth. The frequency shift measured by TDR or microwave techniques is usually analyzed in terms of the relative change in the penetration depth,

$$\Delta f = f(T_{\min}) - f(T) = G\sqrt{\mu(T)}\Delta\lambda_L(T) \qquad (3)$$

where the $\tanh(R/\lambda(\mu))$ term is neglected assuming $\lambda \ll R$ (which is true at low temperatures) and

$$G = \frac{f_0}{R}\frac{V_{sample}}{2V_{coil}} \qquad (4)$$

is the geometric calibration factor that can be measured by pulling the sample out of the coil and calculating the effective dimension $R$ [45]. The $\sqrt{\mu(T)}$ term in Eq. (3) has led to the erroneous misidentification of s-wave pairing in the Nd-based electron-doped superconductors [23, 24] and quite possibly is playing an important role in the iron-based pnictides. With various experimental contrivances, such as active temperature stabilization of the circuit and thermally separating it from the sample, we are able to resolve frequency shifts of about 0.1 Hz at the resonant frequency of $10^7$ Hz. This results in a frequency resolution of 10 ppb, which translates into Angstrom resolution in the London penetration depth.

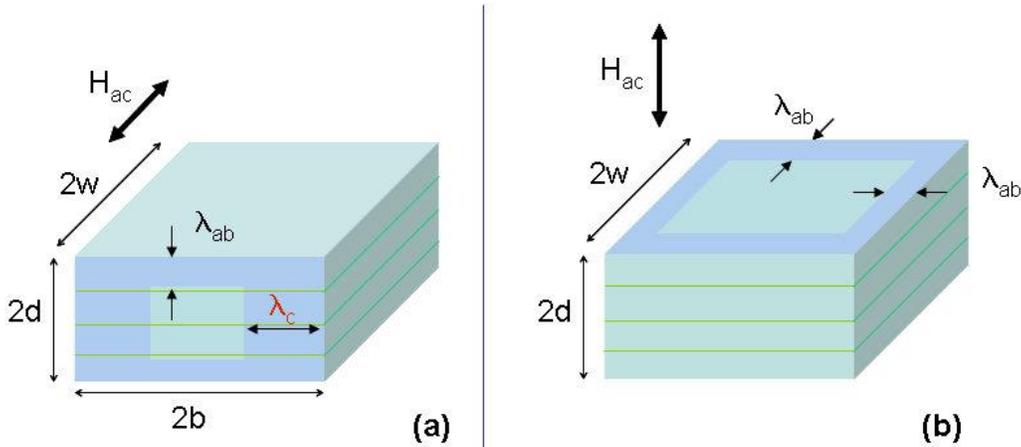

**Figure 2. Experimental configurations. (a) geometry relevant to Eq.(5) when $w \to \infty$. (b) finite $w$ with field normal to conducting planes. In-plane penetration depth is assumed isotropic.**

The anisotropy of the sample leads to additional experimental complications, but can be dealt with by measuring the same sample in different orientations. Let us consider the simplest case with two different London penetration depths, $\lambda_{ab}$ and $\lambda_c$, as shown in Figure 2. For such a geometry, the anisotropic London equation reads [1, 49],

$$\lambda_c^2 \frac{\partial^2 B_z}{\partial x^2} + \lambda_{ab}^2 \frac{\partial^2 B_z}{\partial y^2} = B_z \tag{5}$$

and the overall magnetic susceptibility is given by

$$-4\pi\chi = 1 - \frac{\lambda_{ab}}{d}\tanh\left(\frac{d}{\lambda_{ab}}\right) - 2\lambda_c b^2 \sum_{n=0}^{\infty} \frac{\tanh(\tilde{b}_n/\lambda_c)}{k_n^2 \tilde{b}_n^3} \tag{6}$$

where $k_n = \pi(1/2+n)$ and $\tilde{b}_n = b\sqrt{\left(1+(k_n\lambda_{ab}/d)^2\right)}$. It is easy to show that Eq.(6) gives the correct result for limiting cases. For example, if $\lambda_{ab} \ll d$ and $\lambda_c \ll b$, we obtain $\tilde{b}_n \approx b$ and with $\sum_{n=0}^{\infty} k_n^{-2} = 1/2$, we have $-4\pi\chi = 1 - \lambda_{ab}/d\tanh(d/\lambda_{ab}) - \lambda_c/b\tanh(b/\lambda_c)$, as expected. More often one has $\lambda_c \gg \lambda_{ab}$ and in this case, for typical crystal dimensions, one has $\lambda_c/b \gg \lambda_{ab}/d$, so that the susceptibility is dominated by $\lambda_c$. Nevertheless, by measuring the same sample in different orientations and comparing samples of different aspect ratios, one can extract both $\lambda_{ab}(T)$ and $\lambda_c(T)$. In the following, when we write only $\lambda(T)$, it signifies $\lambda_{ab}(T)$. When we explicitly distinguish the two, it will be indicated.

Another complication arises from the demagnetization factor of the sample. Strictly speaking, it can only be defined for ellipsoidal samples, but for real samples an effective demagnetizing factor is still a useful concept. A more difficult problem is to determine the effective sample dimension by which to normalize the penetration depth in this demagnetizing geometry. A semi-analytical solution for this problem was found in Ref.[21, 45], where it was shown that the magnetic susceptibility is given by an equation similar to that for the case of an infinite slab,

$$-4\pi\chi = \frac{1}{1-N}\left[1-\frac{\lambda}{\tilde{R}}\tanh\left(\frac{\tilde{R}}{\lambda}\right)\right] \quad (7)$$

where $\tilde{R}$ is an effective sample dimension and $N$ is the demagnetization factor. For a disk of thickness $2d$ and radius $w$ and for a magnetic field applied perpendicular to the plane of the disk (i.e. along $d$),

$$\tilde{R} \approx \frac{w}{2\left\{1+\left[1+\left(\frac{2d}{w}\right)^2\right]\arctan\left(\frac{w}{2d}\right)-\frac{2d}{w}\right\}}. \quad (8)$$

In the thin limit, $d \ll w$, $\tilde{R} \approx 0.2w$. For this case, the demagnetization correction is given by,

$$\frac{1}{1-N} \approx 1+\frac{w}{2d}. \quad (9)$$

In Eq. (7), the $\tanh(\tilde{R}/\lambda)$ term is only an approximation (it is exact for an infinitely long slab of thickness $2w$). For an infinitely long cylinder, instead of $\tanh(\tilde{R}/\lambda)$, the exact solution of the London equation gives the ratio of the modified Bessel functions of the first kind, $I_1(\tilde{R}/\lambda)/I_0(\tilde{R}/\lambda)$. However, these distinctions are important only close to $T_c$ (more specifically where $\lambda \geq 0.4w$) and even then the results are quite similar. At low temperatures, $\tilde{R}/\lambda \gg 1$ and the hyperbolic tangent factor is essentially unity and therefore irrelevant. For rectangular slabs Eqs.(7), (8) and (9) can be applied with the effective lateral dimension

$$\tilde{w} = \frac{db}{b+2d/3}. \quad (10)$$

Equation (10) was obtained by fitting the numerical solutions of Eq.(6) in its isotropic form ($\lambda_c = \lambda_{ab}$) to Eq.(7). A straightforward generalization of Eq.(8) would lead to a similar expression, but without the factor of 2/3 in the denominator.

## Results and discussion

### Ba(Fe$_{1-x}$Co$_x$)$_2$As$_2$  (x ≈ 0.038, 0.047, 0.058, 0.074 and 0.10)

This system so far possesses the most uniform and robust superconductivity of all Fe-based pnictides in our experiments, with the results highly reproducible between different batches and agreeing well for different types of measurements.

Figure 3 shows the change in the penetration depth measured for several different doping levels in Ba(Fe$_{1-x}$Co$_x$)$_2$As$_2$ single crystals plotted versus the power-law of reduced temperature, $(T/T_c)^n$, where $n$ was obtained by fitting the data. The values of doping levels and the exponent $n$ are shown in the figure.

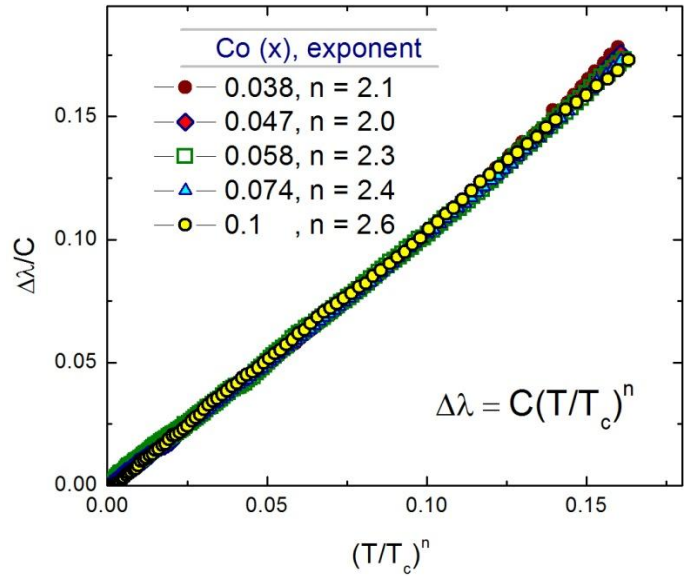

**Figure 3. Relative change in the penetration depth in Ba(Fe$_{1-x}$Co$_x$)$_2$As$_2$ single crystals for indicated x plotted versus normalized power-law behavior.**

Another way to look at our data is to construct the normalized superfluid density,

$$\rho(T) = \left(\frac{\lambda(0)}{\lambda(T)}\right)^2 = \left(1 + \frac{\Delta\lambda(T)}{\lambda(0)}\right)^{-2} \qquad (11)$$

where $\Delta\lambda(T) = \lambda(T) - \lambda(0)$ is the measured change of the penetration depth. The superfluid density is directly related to the superconducting gap structure on the Fermi surface. It is not sensitive to the phase of the order parameter, but it is sensitive to the angular dependence of the gap magnitude and can be used to analyze the unconventional behavior [1, 6].

One problem that exists is that one needs to know the value of $\lambda(0)$. While it is possible to measure it using the TDR technique, this requires additional manipulations of the samples [46] and these results will be reported elsewhere. Here we used measurements of the first critical field, $H_{c1}$, by looking at the first signs of nonlinearity in precision low-field $M(H)$ curves. This procedure works quite well for large samples. Specifically, the sample was cooled in zero field to 5 K (lowest temperature of our SQUID magnetometer, but a lower temperature would be better). After that a small field interval magnetization loop was measured. If the loop was reversible (no noticeable hysteresis), another loop was measured using a higher maximum value of field, $H_{max}$. Results of such measurements are shown in Figure 4 for several values of $H_{max}$. Clearly, the irreversible behavior appears at the field where the deviation from linear behavior is observed. We estimate this field as $65\pm 5$ Oe. We now discuss the procedure used for accounting for the demagnetization factor. By measuring $dM/dH|_{H\to 0}$ and knowing the sample volume we can estimate the effective demagnetization factor, $N=0.64$, which also agrees with the one calculated according to the procedure described in Ref. [1]. We need one more ingredient to estimate $\lambda(0)$, and that is the coherence length, $\xi$. It was measured directly by scanning tunneling spectroscopy to be $\xi=2.76$ nm at 5 K [50]. With this information and using the well-known expression [51]

$$H_{c1} = \frac{\Phi_0}{4\pi\lambda^2}\left(\ln\frac{\lambda}{\xi}+0.485\right) \qquad (12)$$

we estimate $\lambda(5\,\text{K})=210\pm10\,\text{nm}$. By extrapolating the data shown in Figure 3 to $T=0$ we finally obtain $\lambda(0)\approx190\pm10$ nm. (The corresponding Ginzburg-Landau parameter is then $\kappa=\lambda/\xi=76\pm4$). Taking into account the literature data of $\lambda(0)\approx254$ nm for La-1111 [52] and $\lambda(0)\approx190$ nm for Sm-1111 [53] from µSR measurements, we adopt $\lambda(0)=200$ nm for the calculations. The resulting $\rho(T)$ is shown in Figure 5 for two values of $\lambda(0)=200$ nm and 650 nm. The larger value of $\lambda(0)=650$ nm is taken from the theoretical analysis of our data by Wisconsin group [54]. Their explanation invokes the so-called $s_\pm$ pairing model [26, 55] in which non-magnetic impurities act as pair-breakers, similar to magnetic impurities in conventional fully-gapped s-wave superconductors. As shown in Figure 6(a), by tuning the scattering parameters and $\lambda(0)$ it is possible to fit our data quite well, although the values for $\lambda(0)$ are required to be a bit large and the corresponding lower critical field is about 23 Oe. We note, however, that at very low temperatures, the $s_\pm$ scenario will always give exponential behavior, so one of the ways to check its validity is to perform these measurements at very low temperatures using a dilution refrigerator (work in progress). A related result, originally from Ref.[34], is shown in Figure 6(b). Down to 100 mK, the penetration depth is linear in temperature and the $s_\pm$ model cannot fit the data.

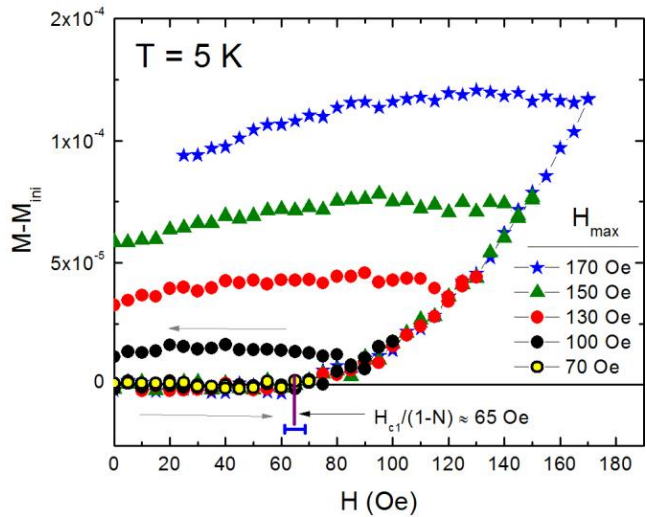

**Figure 4. Partial magnetization loops minus the initial slope at 5 K.**

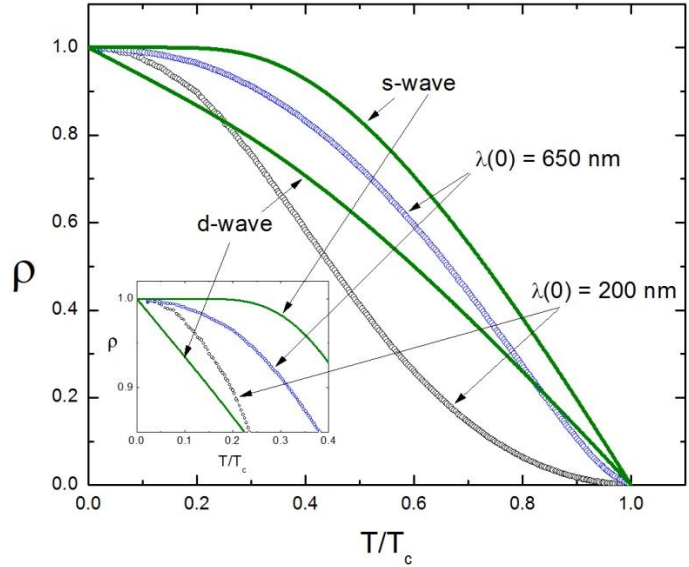

Figure 5. Superfluid density in a Ba(Fe$_{1-x}$Co$_x$)$_2$As$_2$ single crystal near optimal doping of x=0.074. Inset: zoomed in at the lowest temperatures showing the non-exponential behavior. Solid lines show standard weak-coupling s- and d- wave behavior.

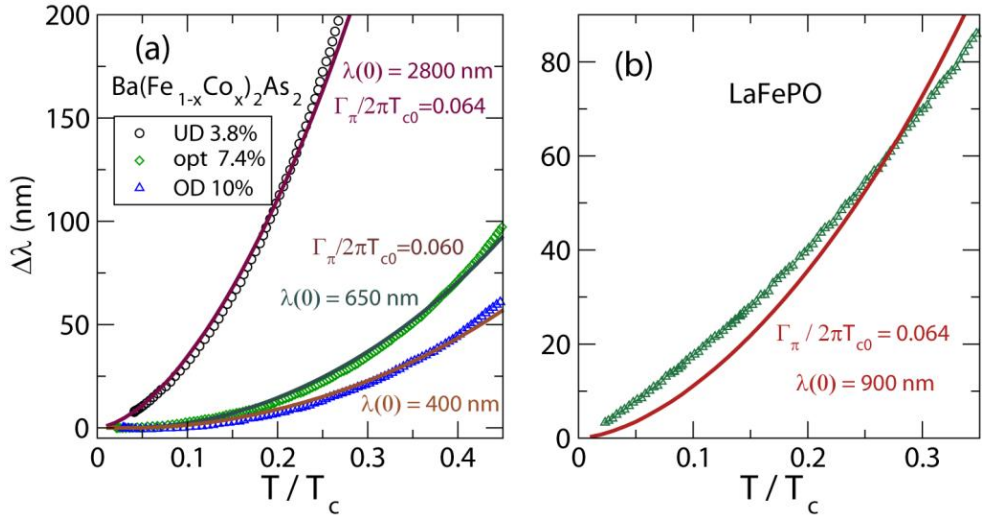

Figure 6. (Reprinted from Ref.[54] by permission from the Authors). (a) Change of the London penetration depth in Ba(Fe$_{1-x}$Co$_x$)$_2$As$_2$ single crystals with x=0.038 (underdoped), x=0.074 (near optimal doping) and x=0.1 (overdoped). (b) $\Delta\lambda(T)$ in clean LaFePO single crystals showing $T-$ linear behavior down to 100 mK [34].

One of the possible reasons for a power-law behavior of the penetration depth, $\Delta\lambda(T) \propto T^n$ with $n \approx 2$, could be that it is a dirty superconductor with line nodes [56]. However, so far, no direct evidence of line nodes exists for the iron-based pnictides. An alternative interpretation proposed in Refs.[31, 32] is the existence of point nodes on three-dimensional sheets of the Fermi surface. In such a case $n \approx 2$ is expected for the clean case and scattering would likely increase this power as observed in our experiments. For a certain doping level such a scenario can be realized for $s_\pm$ pairing with the nodes located between Fermi surface sheets [57]. If the actual shape of the Fermi surface strongly deviates from cylindrical, the equator of the Fermi surface can reach the points where the order parameter changes sign and becomes gapless at a single spot forming a point node. However, such a mechanism depends sensitively on the Fermi surface shape and position (in stark contrast to d- or p-wave states where the nodes are enforced by the symmetry), which in turn depends sensitively on the doping level. Since we have observed a robust and almost universal $\Delta\lambda(T) \propto T^2$ behavior across all doping levels [31], such a scenario needs to be further investigated.

Finally, Figure 7 shows the results of both $\Delta\lambda_{ab}(T)$ and $\Delta\lambda_c(T)$ and the anisotropy ratio, $\gamma_\lambda$. In order to obtain the anisotropy ratio, the value of $\lambda_{ab}(0) = 200$ nm was used and the value for $\lambda_c(0) \approx 1200$ nm was obtained by matching the anisotropy ratio at $T_c$ with the anisotropies of $\gamma_\xi \approx 2$ and $\gamma_\rho \approx 4$ [32, 35]. Contrary to the situation in MgB$_2$ [40], in Ba(Fe$_{1-x}$Co$_x$)$_2$As$_2$ crystals $\gamma_\lambda(T)$ *increases* with the *decrease* of temperature (and $\gamma_\xi(T)$ *decreases* [32]). While an exact treatment is not currently available, it is plausible that such behavior is related to the complex multiband structure of the iron-based pnictides. In MgB$_2$ the anisotropy is dominated by a larger 2D $\sigma$ - band at low temperatures, whereas a more isotropic $\pi$ - band contributes at higher temperatures. In the pnictides such simple arguments may not be valid, especially if one considers inter-band pairing. One conclusion is clear and it is that the electromagnetic anisotropy of the pnictide superconductors is low, in fact much lower than in the high-$T_c$ cuprates.

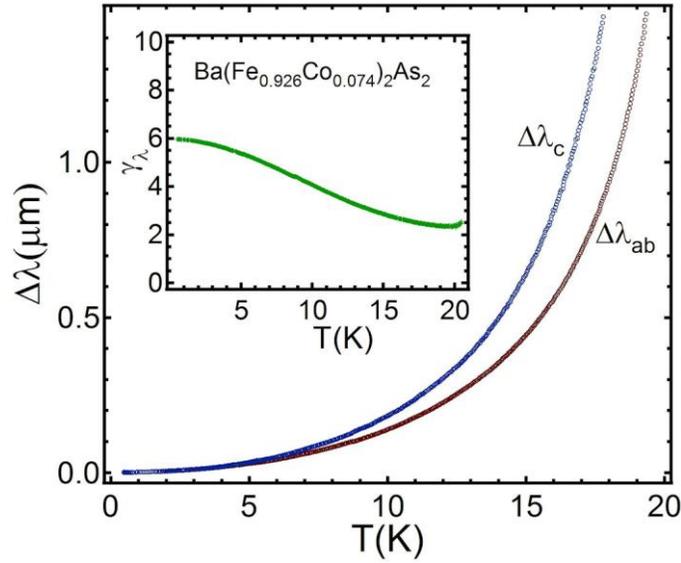

**Figure 7.** Changes in the in-plane and out-of-plane penetration depths in single crystal Ba(Fe$_{1-x}$Co$_x$)$_2$As$_2$ near optimal doping of x=0.074. Inset: anisotropy ratio, $\gamma_\lambda$.

### $(Ba_{1-x}K_x)Fe_2As_2$ ($x \approx 0.45$ and $0.15$)

Several $(Ba_{1-x}K_x)Fe_2As_2$ single crystals of various shapes and aspect ratios, from different batches of Sn-flux grown samples, were studied. A separate report, where crystals grown from both Sn and from FeAs flux were investigated, will be published elsewhere [29] (both show very similar results). The penetration depth for two single crystals is shown in Figure 8. The lower right inset (a) of Figure 8 shows the full temperature range of superconductivity for two samples with nominal potassium content of $x \approx 0.45$ (sample #1 with onset $T_c \approx 28$ K) and underdoped $x \approx 0.15$ (sample #2 with onset $T_c \approx 19$ K). Below $T = 0.3 T_c$, the penetration depth does not show exponential decay as expected from a single superconducting gap with s-wave symmetry. Instead, as it can be seen in the upper left inset (b) of Figure 8, a power-law fit suggests a nearly quadratic behavior of $\lambda(T)$. This behavior is apparent when $\lambda(T)$ is plotted vs. $(T/T_c)^2$, as shown in the main frame of Figure 8. This result implies that the electrodynamic properties are universal, at least within the 122 family. Figure 9 shows the superfluid density in $(Ba_{1-x}K_x)Fe_2As_2$ single crystal #1 with $x=0.45$ for the choice of $\lambda(0)=180$ nm [58, 59]. Obviously, the clean s-wave scenario (double-dot- dashed line in Figure 9) does not reproduce the full temperature dependence of $\rho(T)$. Given the multiband structure of the iron-based pnictide superconductors it is tempting to fit the data to a two-gap model with two s-wave gaps where their amplitudes and the relative contribution of each gap become the free parameters. Such a fit appears to work over the entire temperature scale as shown in Figure 9. Two partial superfluid densities obtained for $\Delta_1 = 23.5$ K and $\Delta_1 = 5$ K are shown as solid lines in Figure 9.

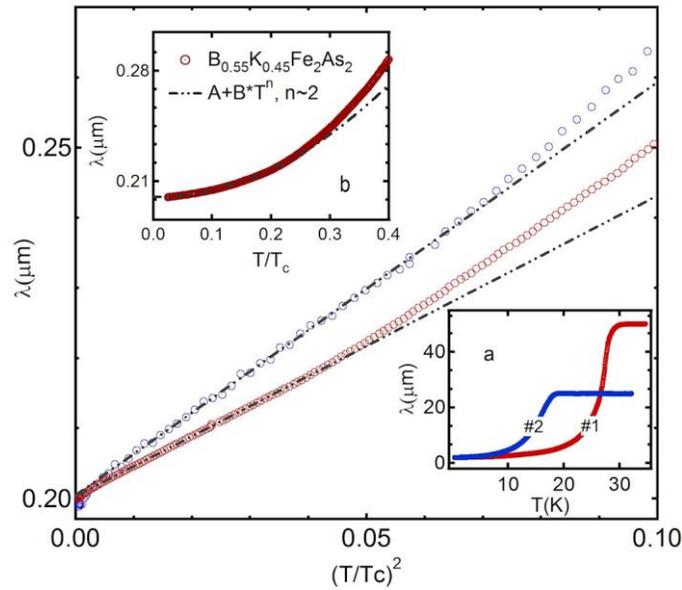

Figure 8. Penetration depth in two different $(Ba_{1-x}K_x)Fe_2As_2$ crystals showing almost perfect $T^2$ behavior at low temperatures. Lower right inset (a) shows the full temperature scale. Upper left inset (b) shows power-law fit.

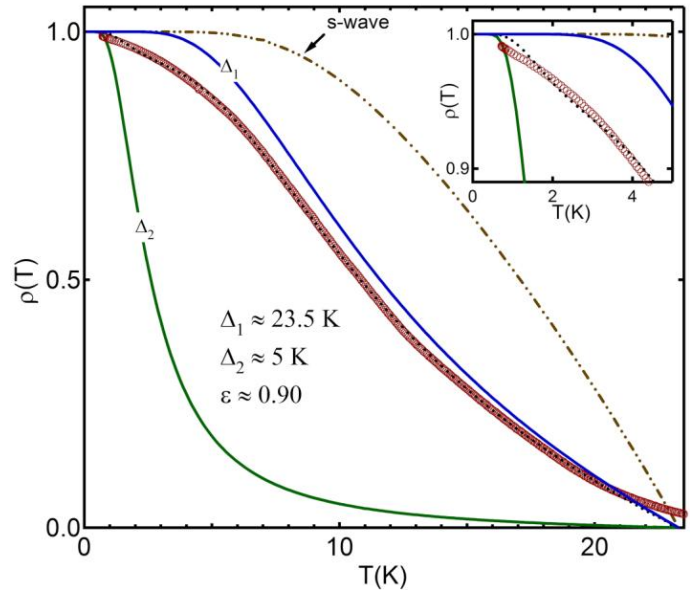

Figure 9. Superfluid density in a $(Ba_{1-x}K_x)Fe_2As_2$ single crystal with $x=0.45$ for $\lambda(0)=180$ nm. Double-dot–dashed line is the standard s-wave curve. Solid lines and a dashed line represent an s-wave two-gap model as explained in the text. Inset: zoomed in at the low-temperature region showing the failure of the two-gap fit.

The resulting superfluid density with 90% contribution from $\Delta_1$ is shown as a dashed line. This fit, however, fails when the low-temperature region is examined as shown in the inset. This does not imply the absence of multi-gap superconductivity but rather it means that the gaps have strong anisotropy and perhaps nodes and/or strong inter-band scattering effects that need to be incorporated into the theoretical models.

Finally, the anisotropy of the penetration depth in a $(Ba_{1-x}K_x)Fe_2As_2$ single crystal with $x = 0.45$ is shown in Figure 10. The procedure used for determining $\gamma_\lambda$ was the same as that described above for the $Ba(Fe_{1-x}Co_x)_2As_2$ crystal and $\gamma_\xi$ was estimated directly from the reported radio-frequency pulsed-field measurements [60]. The overall behavior of both $\gamma_\lambda$ and $\gamma_\xi$ is very close to the Co-doped system, reinforcing the conclusion that these two systems are very similar, despite the difference in carrier type.

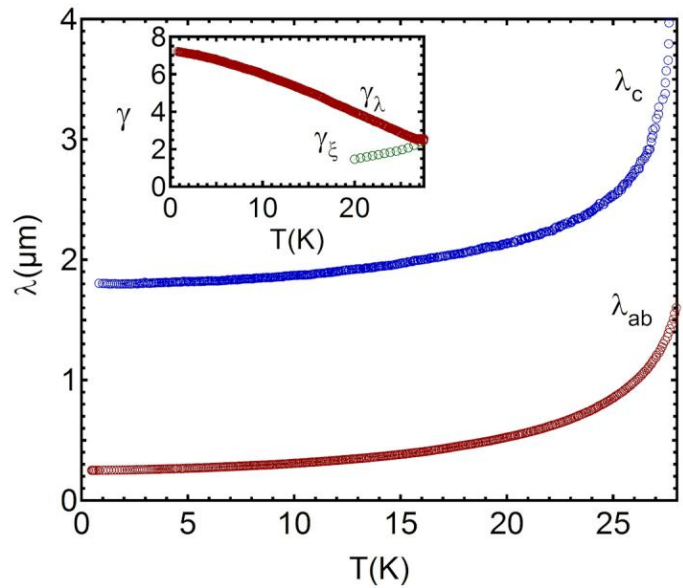

Figure 10. Two London penetration depths in a $(Ba_{1-x}K_x)Fe_2As_2$ single crystal. Inset: anisotropies of the penetration depth, $\gamma_\lambda$, and of the upper critical field, $\gamma_\xi$ [60].

## NdFeAs($O_{1-x}F_x$) (nominal x = 0.1)

The measurements of the NdFeAs($O_{1-x}F_x$) system are more difficult to perform and, at present, the results are more ambiguous due to the small sizes of the crystals and the presence of magnetism due to the rare-earth ions. The crystals were carved out of a dense polycrystalline pellet with the help of magneto-optical mapping [21, 43]. The original samples had the largest dimensions on the order of $100\,\mu m$ and the measurements were quite noisy due to a poor filling factor (ratio of the volumes in Eq.(1)). More recent crystals are 4-5 times larger than the original report [21] and new measurements have given some reason to believe that the conclusion stating that the 1111 system is fully gapped is premature. While these new data are still being analyzed and prepared for publication, here we report on the anisotropy of the penetration depth in NdFeAs($O_{1-x}F_x$) crystals.

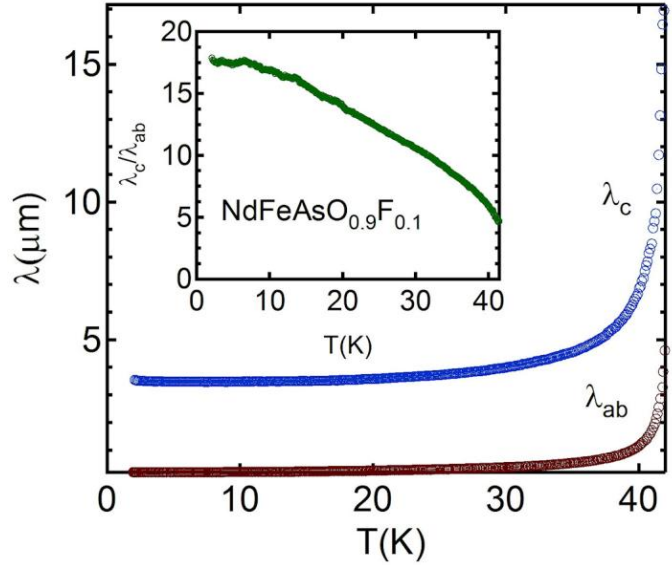

**Figure 11. Penetration depths in NdFeAs($O_{1-x}F_x$) (nominal x=0.1) single crystal. Inset: the anisotropy ratio $\gamma_\lambda = \lambda_c/\lambda_{ab}$.**

Figure 11 shows the London penetration depths in a NdFeAs($O_{0.9}F_{0.1}$) single crystal and the inset shows the anisotropy ratio, $\gamma_\lambda = \lambda_c/\lambda_{ab}$. Clearly, this material is more anisotropic, so that $\gamma_\lambda$ changes from about 4 at $T_c$ to about 18 at low temperatures. Similar trend was observed in Nd-1111 and Sm-1111 crystals by torque measurements [37-39]. For comparison, the anisotropy changes from 2 to 7 in BaK-122 and from 2 to 6 in FeCo-122, so it is about two times more anisotropic than the 122 pnictides. In either case, the temperature dependence of $\gamma_\lambda(T)$ remains similar in that it increases upon the decrease of temperature.

# Conclusions

In conclusion, we have studied several dozens of single crystals of three families of iron-based pnictide superconductors and from different growth procedures. While the behavior of the 1111 system is still an open question due to small crystal sizes and magnetism of rare-earth ions, the London penetration depth is clearly non-exponential in all 122 samples. For both hole- and electron-doped compounds and for all doping levels from the underdoped to overdoped regimes, the penetration depth exhibits a robust and almost universal power-law behavior, $\lambda(T) \propto T^n$ with the exponent $n \approx 2-2.5$, depending on the studied samples. Three primary mechanisms that have been suggested to explain such behavior are: 1) dirty superconductor with line nodes; 2) dirty superconductor within $s_\pm$ pairing scenario; 3) superconductor with point nodes. At this stage we cannot give priority to any of these mechanisms. However, the universality of the behavior (i.e., the lack of significant dependence on the doping) suggests that the results are unrelated to impurity scattering. Perhaps yet another unknown mechanism is at play and this paper should be considered as a summary of a comprehensive experimental study of the penetration depth in single crystals rather than an attempt to promote any specific pairing model.

We have also reported direct measurements of the electromagnetic anisotropy, $\gamma_\lambda = \lambda_c/\lambda_{ab}$, in all studied systems. The anisotropy of the 1111 system is about twice that of the 122 system, but both remain quite low compared to the high-$T_c$ cuprates. It is possible that the low anisotropy provides additional support for the existence of a fairly three-dimensional character to the electronic structure of the pnictide superconductors making the point-node scenario more plausible.

# Acknowledgements

We thank A. Carrington, A. V. Chubukov, J. R. Clem, P. J. Hirschfeld, A. Kaminski, I. I. Mazin, G. D. Samolyuk and J. Schmalian for discussions and comments. We thank A. V. Chubukov and A. B. Vorontsov for the permission to use Figure 6 and fitting our data. Work at the Ames Laboratory was supported by the Department of Energy-Basic Energy Sciences under Contract No. DE-AC02-07CH11358. R. P. acknowledges support from Alfred P. Sloan Foundation. M.A.T. acknowledges continuing cross-appointment with Institute of Surface Chemistry, NAS Ukraine.